\documentclass[aps,reprint,superscriptaddress,longbibliography]{revtex4-1}

\usepackage{setspace}

\usepackage{hyperref}
\usepackage{outlines}
\usepackage{fancyvrb}
\usepackage{amsmath}
\usepackage{hyperref}
\usepackage{braket}
\usepackage{comment}
\usepackage{bbold}
\usepackage{amssymb}
\usepackage{graphicx}
\usepackage[caption=false]{subfig}
\usepackage[usenames, dvipsnames]{color}
\usepackage{soul}
\usepackage{amsthm}

\newtheorem{lemma}{Lemma}[section]

\begin{document}


\title{Learning Simon's quantum algorithm}


\author{Kwok Ho Wan}
\affiliation{Mathematical Physics, Department of Mathematics,
  Imperial College London, London, SW7 2AZ, United Kingdom}
  \affiliation{QOLS, Blackett Laboratory,
  Imperial College London, London, SW7 2AZ, United Kingdom}

\author{Feiyang Liu}
\affiliation{Institute for Quantum Science and Engineering, Department of Physics, South China University of Science and Technology (SUSTech),  Shenzhen, China.}

\author{Oscar Dahlsten }
\thanks{Correspondence: Oscar Dahlsten \\(dahlsten@sustc.edu.cn)}
\affiliation{Institute for Quantum Science and Engineering, Department of Physics, South China University of Science and Technology (SUSTech),  Shenzhen, China.}

\affiliation{QOLS, Blackett Laboratory,
  Imperial College London, London, SW7 2AZ, United Kingdom}
\affiliation{Clarendon Laboratory,
  University of Oxford, Parks Road, Oxford, OX1 3PU, United Kingdom}
\affiliation{London Institute for Mathematical Sciences, 35a South
  Street Mayfair, London, W1K 2XF, United Kingdom}
\author{M.S. Kim}
\affiliation{QOLS, Blackett Laboratory,
  Imperial College London, London, SW7 2AZ, United Kingdom}



\date{\today}

\begin{abstract}
We consider whether trainable quantum unitaries 
can be used to discover quantum speed-ups for classical problems. Using methods recently developed for training quantum neural nets, we consider Simon's problem, for which there is a known quantum algorithm which performs exponentially faster in the number of bits, relative to the best known classical algorithm. We give the problem to a randomly chosen but trainable unitary circuit, and find that the training recovers Simon's algorithm as hoped.
\end{abstract}


\maketitle


\section*{Introduction}
\textit{The power of quantum computation} by~\cite{simonoriginalpaper}
Simon provides an exponentially faster quantum algorithm compared to a classical randomised search. Simon illustrated a very simple and scalable quantum circuit to solve a mathematical game now known as Simon's problem. The aim is to learn a property of a black-box function, a secret bit string $s$, which determines the function within the families of functions under consideration. Simon's quantum algorithm is an important precursor to Shor's Algorithm \cite{Shor:1997:PAP:264393.264406}, which also provides an
exponential speed up over the best known classical algorithms. With a quantum computer, one could employ Shor's algorithm to quickly break the widely used RSA cryptographic protocol \cite{mermin2007quantum}. Simon's and Shor's algorithms are both examples of the Hidden Subgroup Problem over Abelian groups~\cite{Grigni:2001:QMA:380752.380769}.

 Quantum machine learning, see e.g.~\cite{Biamonte, Schuld14, LloydMR13, LloydMR13ii, Montanaro15, Aaronson15, GarneroneZL12, HarrowHL09, LloydGZ16, RebenstrostML13, WiebeBL12, Adcock15, HeimRIT15, GrossYFBE10, Dunjko16, Wittek14}, contains a  research direction known as {\em
  quantum learning} ~\cite{Bisio10, Sasaki02, BangLKL08, Sentis15, Banchi16, Palittapongarnpim16, WanDKGK17} which concerns learning and optimising with truly quantum objects. In~\cite{WanDKGK17} some of the present authors defined a quantum generalisation of feedforward neural networks which could numerically be trained to perform various quantum generalisations of classical tasks. This motivated us to consider whether these networks can also find quantum speed-ups for classical tasks. This could help deal with the shortage of useful quantum algorithms. To test this, Simon's algorithm is a natural candidate, having an exponential speed-up over the best known classical algorithm at the same time as being a more minimal, and thus more tractable, algorithm than Shor's.
  
We here accordingly aim to determine whether a quantum neural net can discover Simon's algorithm.
We design an explicit training procedure, and demonstrate that it works. This gives significant hope that it is possible to discover new algorithms using this method of quantum learning.


\section{Technical Intro}
The notation: $\ket{a}^{\otimes N} \equiv
\overbrace{\ket{a} \otimes \ket{a} \otimes ... \otimes \ket{a}}^{N \ of
  \ these \ \ket{a}}$,
will be used throughout. Note that the words ``gates" and ``unitaries"
will be used synonymously throughout. Also, the words ``blackbox
function" and ``oracle" are interchangable.
\subsection{Simon's algorithm}
Simon's problem and solution can be summarised as follows~\cite{simonoriginalpaper, vaziranilec7}. There is a blackbox function, or oracle, that holds a secret string, $s$, within it. One can ask the oracle questions by
querying it. The goal is to infer the secret string $s$ with
the least number of queries. This blackbox function could be represented classically as
$f(x)$ - a function that takes an $n$ bitstring, $x =
x_{1} x_{2} x_{3}...x_{n}$, as an input, where $x_{i}$ is either
zero or one. The $n$ bitstring, $x$, lives in the set, $\{0,1\}^{n}$,
which is the collection of all possible $n$ bitstrings. $f(x)$ is by design guaranteed to either be a particular type of many-to-one functions or a one-to-one function. We restrict, for simplicity, $f(x)$ further by excluding the one-to-one case. 

\begin{figure}[ht]
\includegraphics[width=\linewidth]{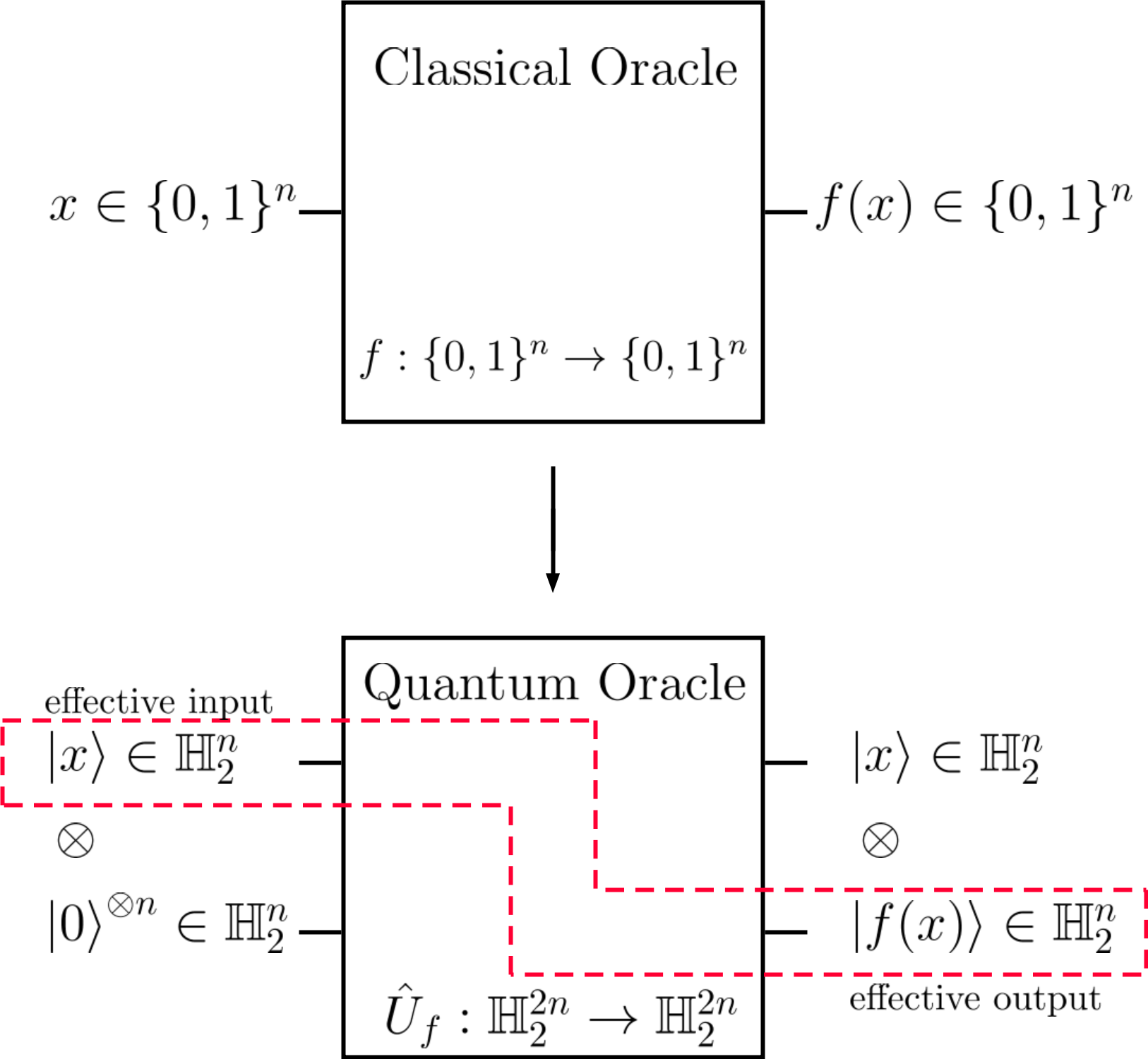}
\caption{\label{fig:generalisation_cl_qm} This shows the
  generalisation of a classical function oracle to a quantum unitary
  operator oracle, which are maps that perform: $f:\{0,1\}^{n}
  \rightarrow \{0,1\}^{n}$ and $\hat{U}_{f} : \mathbb{H}_{2}^{2n}
  \rightarrow \mathbb{H}_{2}^{2n}$ respectively. Note that the qubit lives in the space: $\mathbb{H}_{2}$. The red box in the
  lower quantum circuit diagram shows the effective input to output of the
  quantum string as the generalisation of the oracle to a quantum one
  is reversible.}
\end{figure}

In the quantum version, the blackbox function generalises to a unitary transformation of states: $\hat{U}_{f} \in
\mathbb{H}_{2}^{\otimes 2n}$. In Simon's solution, the quantum state:
$\ket{x} = \ket{x_{1}}\otimes \ket{x_{2}}\otimes ... \otimes \ket{x_{n}} \otimes
\ket{0}^{\otimes n}$ encodes the same bit string $x =
x_{1} x_{2} x_{3} ... x_{n}$. The classical to quantum generalisation
of the oracle can be pictured through Fig.~\ref{fig:generalisation_cl_qm}.
The quantum version of the oracle acts on a quantum bitstring of length $2n$ instead of $n$. This is a well known technique in quantum
generalisations. Since closed quantum evolution is inherently unitary
(reversible), we need the oracle to map the initial state, $\ket{x}
\otimes \ket{0}^{\otimes n}$, to the final state $\ket{x} 
\otimes \ket{f(x)}$ in order to make the map reversible. The extra
padding of qubits ($\ket{0}^{\otimes n}$ after the effective output) allows the back-tracking of the input given the output,
hence the map $\hat{U}_{f}$ is reversible.
In summary, Simon's problem is as follows.

\begin{enumerate}
\item $s$ is an $n$ bit string.
\item Blackbox function $f :\{0,1\}^{n} \rightarrow \{0,1\}^{n}$
\item $\exists$ secret string $s \in \{0,1\}^{n} \neq \overbrace{00...000}^{n}$ such that for all
inputs $x \in \{0,1\}^{n}$ , $f(x) = f(x\oplus s)$
\item For all inputs $x, y \in \{0,1\}^{n}$, if $x \neq y\oplus s \Rightarrow
f(x) \neq f(y)$,
\end{enumerate}
where $\oplus$ is the bitwise modulo 2 addition of two $n$
bitstrings. The game is to find the secret string $s$. Note that the modulo 2 bitwise addition of $s$ to any input $x$ will not change the outcome of the function. 

\begin{figure}[ht]
\includegraphics[width=\linewidth]{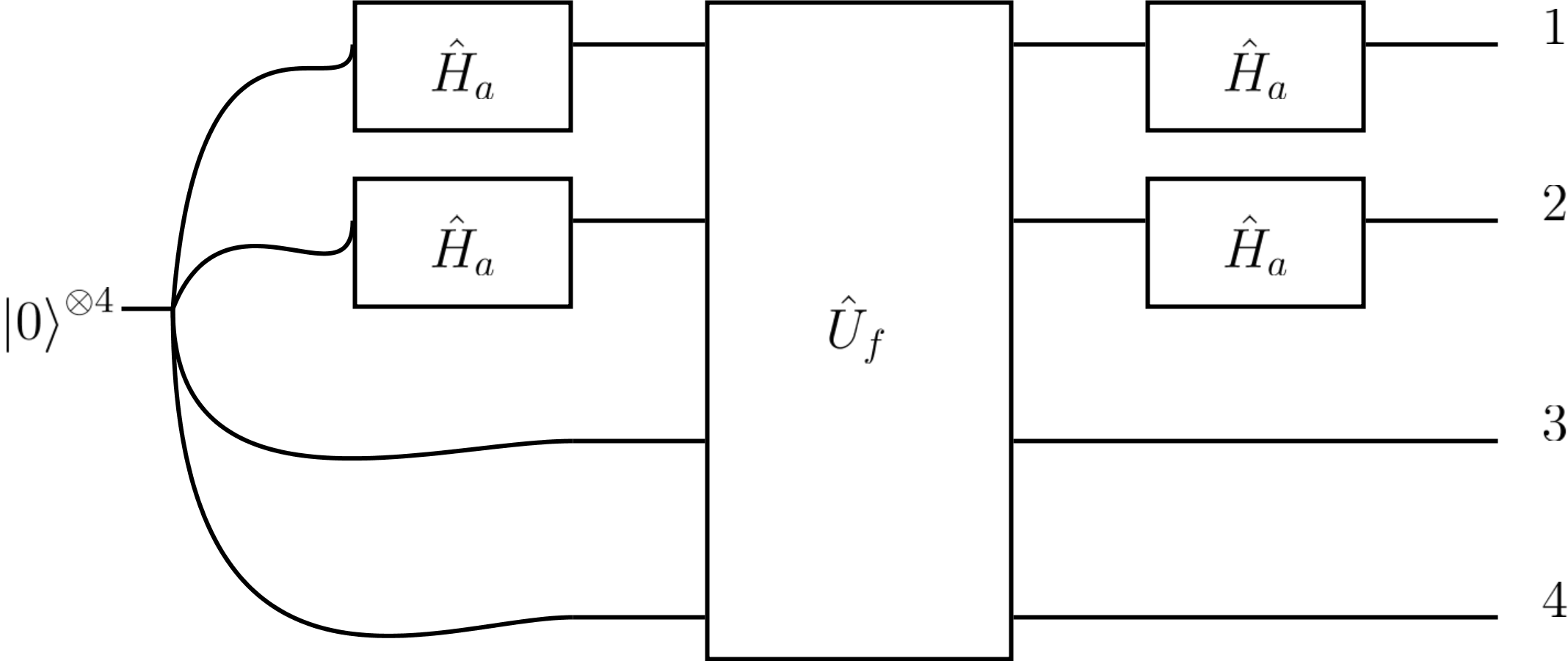}
\caption{\label{fig:n_2_simon_circuit} Simon's quantum circuit for $n = 2$. The 2 top qubits are associated with the input to the function and the two lower to the output. Hadamard unitaries create a superposition of inputs. $U_f$ enacts the function reversibly, storing the input in the upper two qubits. 
Hadamards again after $U_f$ create further input superposition branching, allowing for interference between terms which originally had different inputs and the same outputs.
The number on the right-hand side labels
  the output port number.}
\end{figure}

In Simon's quantum scheme, one has to have access to a $2n$ bitstring
long quantum state, $\ket{0}^{\otimes 2n}$ in the computational basis.
One applies the gate
$\hat{U}_{Simon} = \hat{H}_{a}^{\otimes n}\otimes\mathbb{1}_{2}^{\otimes n}$ to the initial
state $\ket{0}^{\otimes 2n}$ before and after the application of the
fixed secret string
oracle unitary, $\hat{U}_f$ as shown in Fig.~\ref{fig:n_2_simon_circuit}. The matrix $\hat{H}_{a}$ is the 2 by 2 Hadamard
matrix and $\mathbb{1}_{2}$ is the 2 by 2 identity matrix. In the
basis of the Pauli-$z$ eigenstates $\{ \ket{0}, \ket{1} \}$, $\hat{H}_{a} =
\frac{1}{\sqrt{2}} \begin{bmatrix}
   1 &1\\
   1 &-1\\
\end{bmatrix}$.
Hence the final state from the quantum circuit
is: 
\begin{equation}
\label{eq:simon_original_soln}
\ket{final} = \hat{U}_{Simon} \cdot \hat{U}_{f} \cdot \hat{U}_{Simon}
\ket{0}^{\otimes 2n} \ \ .
\end{equation}
This can be represented using the quantum circuit diagram in Fig.~\ref{fig:n_2_simon_circuit} for $n = 2$.

We shall represent the secret string as: 
\begin{equation}
\label{eq:s_string}
s = s_{1}s_{2}...s_{n-1}s_{n} \ \ ,
\end{equation}
where $s_{i} \in \{0,1\}$ is the $i^{th}$ bit in the $n$
bit string $s$.
\\\\
In Simon's Algorithm, one measures the first register (output port $1$
to $2$ in Fig.~\ref{fig:n_2_simon_circuit}). This will produce an $n$ bit string, $y$, such that the dot
product between $y$ and $s$ in $\mod  2$ is zero, i.e. $y \cdot s = 0 \pmod{2}$. The algorithm requires repeated inquiries to the oracle,
hence obtaining many different $n$ bit strings, $y^{(i)}$, with $i$ indexing the results obtained from each inquiry. The $y$ obtained will be $\{y^{(1)} , y^{(2)}, y^{(3)}, ... ,
y^{(J)}\}$, for $J$ inquiries. Then a
classical processing task of Gaussian Elimination in the Galois field $GF(2)$ is carried out to find $s$, represented as follows:

\begin{equation}
\begin{bmatrix}
   0 &0& ... & 0\\
    y_{1}^{(1)} &y_{2}^{(1)} & ... & y_{n}^{(1)}\\
    \cdot     &\cdot&&\\
    \cdot     &&\cdot&\\
    \cdot     &&&\cdot\\
    y_{1}^{(n-1)} &y_{2}^{(n-1)} & ... & y_{n}^{(n-1)}\\
\end{bmatrix}
\begin{bmatrix}
    s_{1}\\
    s_{2}\\
    \cdot\\
    \cdot\\
    \cdot\\
    s_{n}\\
\end{bmatrix}
=
\begin{bmatrix}
0\\
0\\
0\\
0\\
0\\
0\\    
\end{bmatrix} \pmod{2} \ \ .
\end{equation}
Solving the linear equations in $GF(2)$ is equivalent to only
permitting the $s$ vector's elements to be in $\{ 0, 1 \}$, solving the equations
in modulo 2.
\\\\
Since we require that $s$ is not the zero string, only $n-1$
linearly independent equations are needed. The measurement process is probabilistic, meaning sometimes one might not get a set of linearly independent
equations to perform Gaussian elimination on, hence one has to inquire the oracle more than $n-1$ times to get a unique solution for
$s$, which means $J > n-1$ on average. The Gaussian elimination would have at worst $O(n^{2.3755})$ complexity
in time overhead, because the fastest classical algorithm to solve
linear equations by
Coppersmith and Winograd \cite{cwalgorithm}, scales in that manner. 

\subsection{General Unitary Matrix Parameterisation} We shall train over families of unitaries using techniques from~\cite{WanDKGK17}. We shall use a general form of a unitary
matrix in terms of Pauli matrices. A general 1 qubit unitary circuit could be written in the forms: 
\begin{equation}
\label{eq:general_1_qubit_unitary}
\begin{split}
\hat{U}^{(1 \ qubit)} & = \exp\Bigg(i \sum_{j = 0}^{3}\alpha_{j}\hat{\sigma}_{j}\Bigg), \text{or} \\
\hat{U}^{(1 \ qubit)} & = e^{i \alpha_{0}} \Bigg( \cos
\Omega ~\mathbb{1} + i
\frac{\sin{\Omega}}{\Omega} \sum_{j=1}^{3} \alpha_{j} \hat{\sigma}_{j}
\Bigg),
\end{split}
\end{equation}
where $\{ \hat{\sigma}_{0} , \hat{\sigma}_{1} , \hat{\sigma}_{2} , \hat{\sigma}_{3} \}$ are
the 2 x 2 identity, Pauli-$x$, $y$ and $z$ matrices respectively
and $\alpha_{i} \in [0,2\pi)$ and $\Omega  = \sqrt{\alpha_{1}^{2}+\alpha_{2}^{2}+\alpha_{3}^{2}}$.

A useful special case with one parameter we will use is 
$U=\cos(\theta)\hat{\sigma}_{3}+\sin(\theta)\hat{\sigma}_{1}$. A general two qubit unitary could be written in a similar form: 
\begin{equation}
\label{eq:general_2_qubit_unitary}
\begin{split}
\hat{U}^{(2 \ qubits)} & = \exp \Bigg(i \sum_{j=0}^{3}\sum_{k =
  0}^{3}\alpha_{j,k}(\hat{\sigma}_{j} \otimes \hat{\sigma}_{k}) \Bigg)
\ \ .
\end{split}
\end{equation}
\section{Methods}
\subsection{Overall approach}
\begin{figure}[ht]
\includegraphics[width=\linewidth]{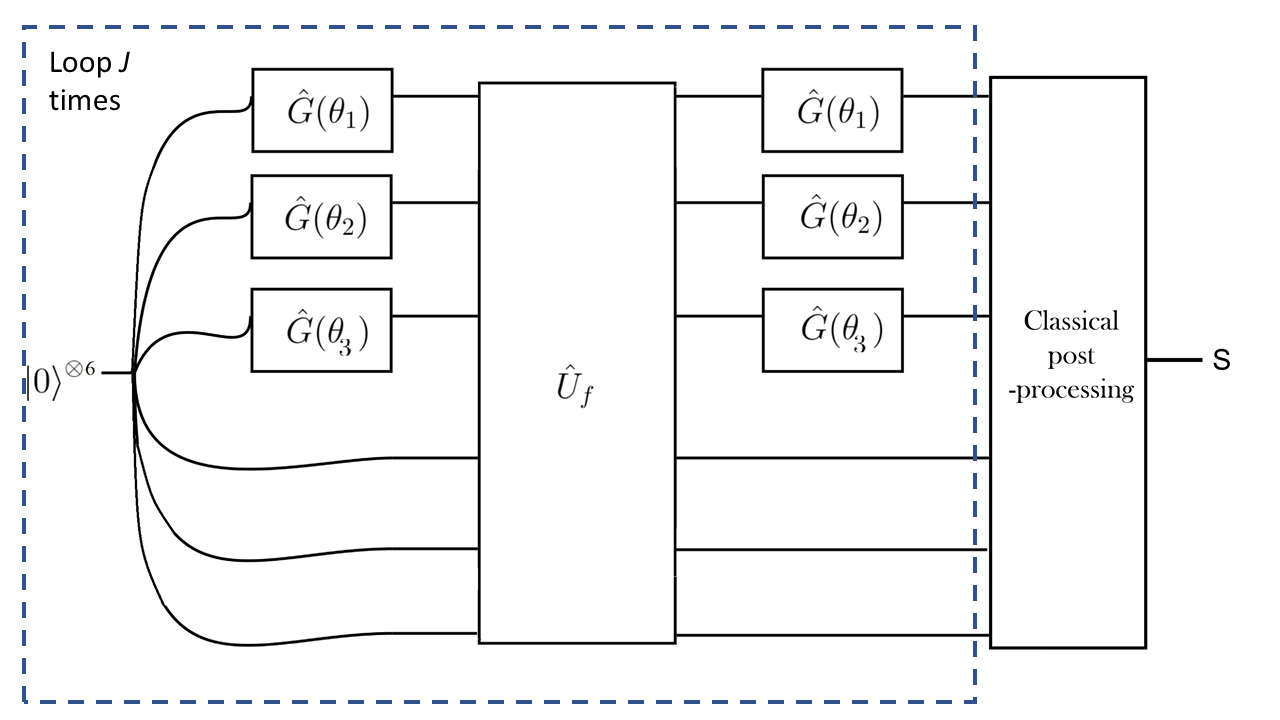}
\caption{\label{fig:recovery_circuit} Gradient descent is performed on free parameters 
  $\theta_{1}$, $\theta_{2}$ and $\theta_{3}$,  with respect to a single
  cost function, over all training examples. The quantum part is looped over $J>n-1$ times in line with the technical introduction. }
\end{figure}

The overall approach is described in Fig.~\ref{fig:recovery_circuit}. The black-box unitary is given, and the quantum circuit together with the classical post processing needs to learn which states to inject and what type of post-processing to do. There are single-qubit unitaries (or 2 qubit unitaries) with free parameters which get tuned in a systematic manner during the training procedure. There is also a classical post-processing part, which could in principle be a quantum circuit but the number of qubits required would be impractical for simulation/experiments, so it is kept classical.  
%
%
%

%
%


%
\subsection{Number of possible training examples}

\begin{lemma} 
\label{lemma_1}
The number of 2 to 1 functions of $n$ bitstrings is
$^{2^{n}}C_{2^{n-1}}$ for a given $s$.
\begin{proof}
There are $2^n$ possible $n$ bitstrings. The statement of the lemma is
equivalent to saying how many different
ways are there to pick $\frac{2^n}{2}$ objects from a set of $2^{n}$
objects disregarding ordering. \\\\ 
$\Rightarrow \ ^{2^{n}}C_{2^{n-1}}$
different 2 to 1 functions of $n$ bitstrings.
\end{proof}
\end{lemma}
\begin{lemma} 
For each $n\in \mathbb{Z}$ in Simon's Problem, there are $(2^{n} - 1) \cdot
(^{2^{n}}C_{2^{n-1}})$ possible mapping tables. 
\begin{proof}
The oracle can have $^{2^{n}}C_{2^{n-1}}$ different functions for
a given $s$ (using lemma \ref{lemma_1}) and there are $(2^{n} - 1)$ different possible $s$ as the
problem excludes the zero $n$ bit-string. 
\\\\
$\Rightarrow (2^{n} - 1) \cdot (^{2^{n}}C_{2^{n-1}})$ is the number of
mapping tables for a specific $n$ in Simon's Problem.
\end{proof}
\end{lemma}
%
\subsection{Cost function}
The cost function, which is the mathematical representation of the aim of the task, is
\begin{equation}
\label{eq:costfunction}
\begin{split}
C & =   \sum_{\forall s \neq {00...000}} (p_{desired}^{s}-p^{s})^2,  
\end{split}
\end{equation}
where $p_{desired}^{s}$ is the probability of the output bit string we want it to be and  $p^{s}$ is the probability of the measured value. The value of $C$ depends on the free parameters being tuned. 
This is a supervised learning scenario as the correct $s$ is known and used to evaluate the cost function~\cite{Nielsen_nn_online_bk}.
\subsection{Gradient descent}
 The quantum circuit is systematically tuned until it
reaches a minimum turning point in the cost function. This could be
achieved through gradient descent with respect to the $\alpha_{j,k}$ parameters
from the unitaries in Eq.~\ref{eq:general_1_qubit_unitary}. Gradient descent is defined as~\cite{Nielsen_nn_online_bk}: 
\begin{equation}
\alpha^{(l)}_{j,k} \rightarrow \alpha^{ (l) }_{j,k} - \eta \frac{\partial
  C}{\partial \alpha^{ (l)}_{j,k}}, \ \forall j,k \ \ ,
\end{equation} where $\eta \in \mathbb{R}^+$ is the step size
of the gradient descent and $l$ labels the different unitaries.

\subsection{Gradient Descent Assisted Genetic Algorithm
Search}
Whilst gradient descent works well in many examples, it is a reasonable assumption that optimisation in a high dimensional parameter
space may have many local minima. In order to get out of a local
minimum \cite{gene_1}, a Genetic Algorithm is used in the
optimisation. This
could be done in parallel, which is worthwhile when the computation
is done in a supercomputer or using GPUs with a multitude of cores. The gradient assisted genetic algorithm implemented in my system was as
follows heuristically:
\begin{enumerate}
\item An agent is a random initial guess for
  the parameters of the unitaries. Start with many agents. This will involve simultaneously
initialising many different sets of $\alpha_{j,k}^{(l)}$, with each
set being a different agent. The genetic information of
each agent is the $\alpha_{j,k}^{(l)}$ parameters.
\item Carry out gradient descent on each agent individually. In this
parallelisable procedure, gradient descent is performed for a small
number of steps. We shall call this number of steps in gradient descent
a generation.
\item After one generation, compare the cost function of each of the
agents and find the $\alpha_{j,k}^{(l)}$ of the few agents with the lowest
cost. Then repopulate the entire population with the selected few
agents with the lowest cost.
\item Apply a small probabilistic random parameter to the $\alpha_{j,k}^{(l)}$ while repopulating
the population. This would be analogous to the mutation process in
Biology.
\item Repeat the procedure until a minimum is found.
\end{enumerate}

\begin{figure}[ht]
\includegraphics[width=\linewidth]{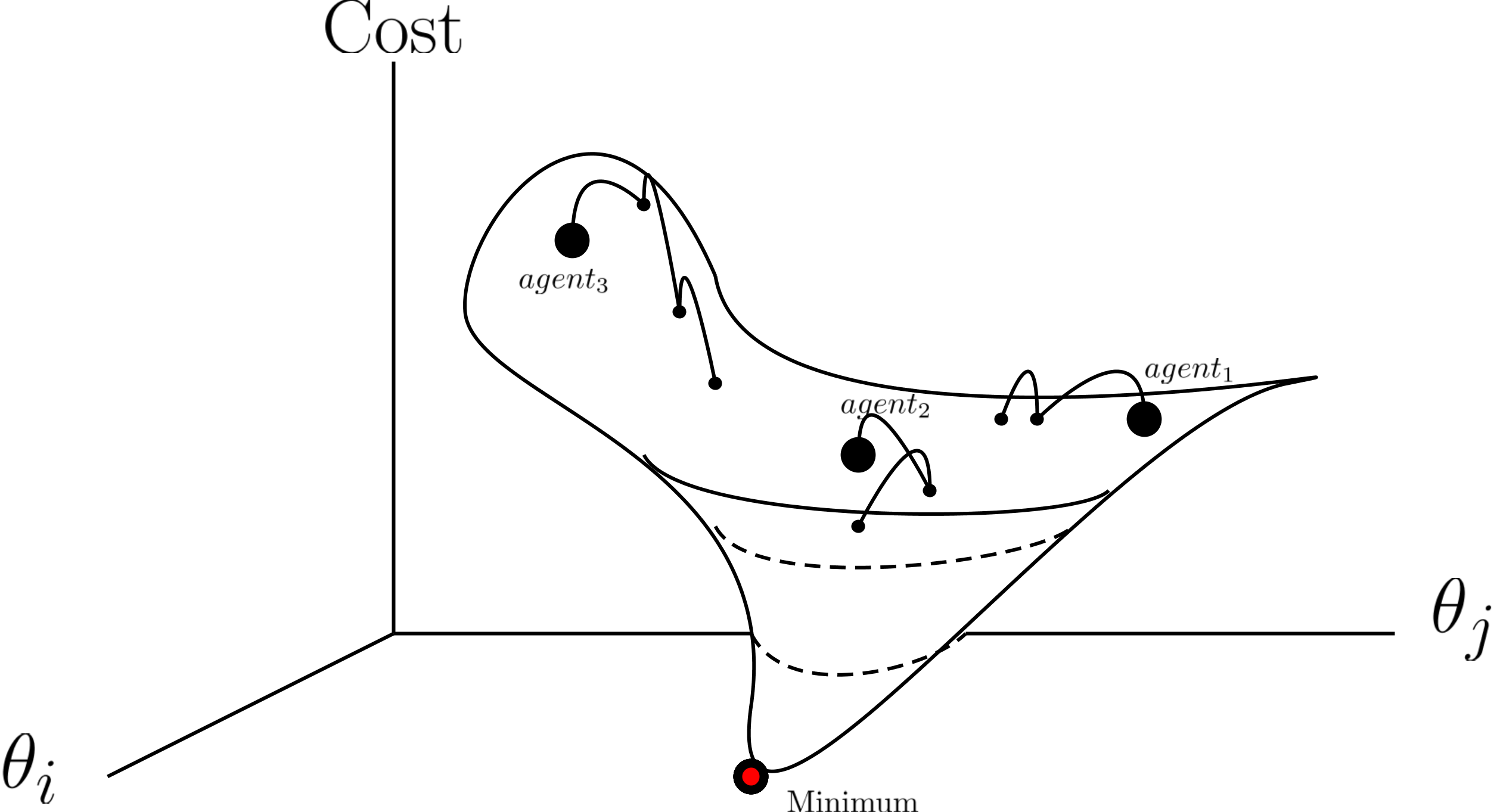}
\caption{\label{fig:genetic_search} A pictorial representation
  of how a genetic algorithm works. Many agents are initialised and
  they are all allowed to propagate to find the minimum.}
\end{figure}

Despite the high computational cost, this method will not guarantee that the final solution will be a global minimum. However, the benefits of this type of search is that the algorithm
is now very parallelisable and the mutations added may aid the agents
in getting out of a local minima. See Fig.~\ref{fig:genetic_search} for a pictorial
description of the algorithm.

\subsection{Classical post-processing}
The classical post-processing takes a set of outputs (the classical bit string $y$'s) and maps them to the corresponding guess for the secret bit-string: $s$. It thus plays an essential role in the algorithm. In Simon's algorithm this is done by Gaussian elimination modulo 2, as discussed in the technical introduction. 

 Whilst this part could in principle be enacted with a quantum circuit, as quantum unitary circuits generalise classical computing, that is very costly experimentally and in terms of classical numerical simulation. Our approach here is thus, 
as depicted in Fig.~\ref{fig:recovery_circuit}, to include classical post-processing, which takes classical input(s) and then gives the answer: the secret bit string $s$. As Simon's algorithm  and other similar algorithms require several outputs before the classical post-processing, we loop over the quantum part $J$ times to give $J$ classical outputs which are then fed to the classical post-processing. 

The classical post-processing amounts to a classical input-output function. This function can be trained as part of the overall training or to make it simpler it can be set by hand to what we want it to be. In the case where only one $s$ was shown to be needed for $n=2$ and $n=3$, the table was set by hand. For $n=2$ we have also tested that the classical part can be trained together with the quantum part rather than set by hand. The training of this part was done by switching one uniformly randomly chosen pair of output bit strings at a time, and accepting the switch only if it decreased the cost function.

\section{Results}
The main results are summarised as follows.
\subsection{Recovering Simon's Circuit}
\newtheorem*{theorem}{Results} 
\begin{theorem}
A set of restricted unitaries in the form of
\begin{equation}
\label{eq:restricted_unitary}
\hat{G}(\theta_{i})= 
\begin{bmatrix}
    \cos(\theta_{i}) & \sin(\theta_{i})    \\
    \sin(\theta_{i})  & -\cos(\theta_{i}) 
\end{bmatrix}
\end{equation}
could be put in place of the Hadamard matrices in Simon's original circuit, as shown in Fig.~\ref{fig:recovery_circuit}, such that gradient descent could be
performed on $\theta_{1}$, $\theta_{2}$ and $\theta_{3}$ with respect to the cost function of Eq.~\ref{eq:costfunction} to
recover Simon's circuit. 
\end{theorem}

Starting with general single-qubit unitaries yields the same performance as in that restricted single qubit unitary case.
\begin{theorem}
The same training performed on a set of restricted unitaries in the form of Eq.~\ref{eq:general_1_qubit_unitary} recovers a circuit with the same performance (same cost function minimum) as Simon's circuit, but not necessarily the Hadamard gates as in Simon's circuit.
\end{theorem}

\begin{figure}[ht]
\includegraphics[width=\linewidth]{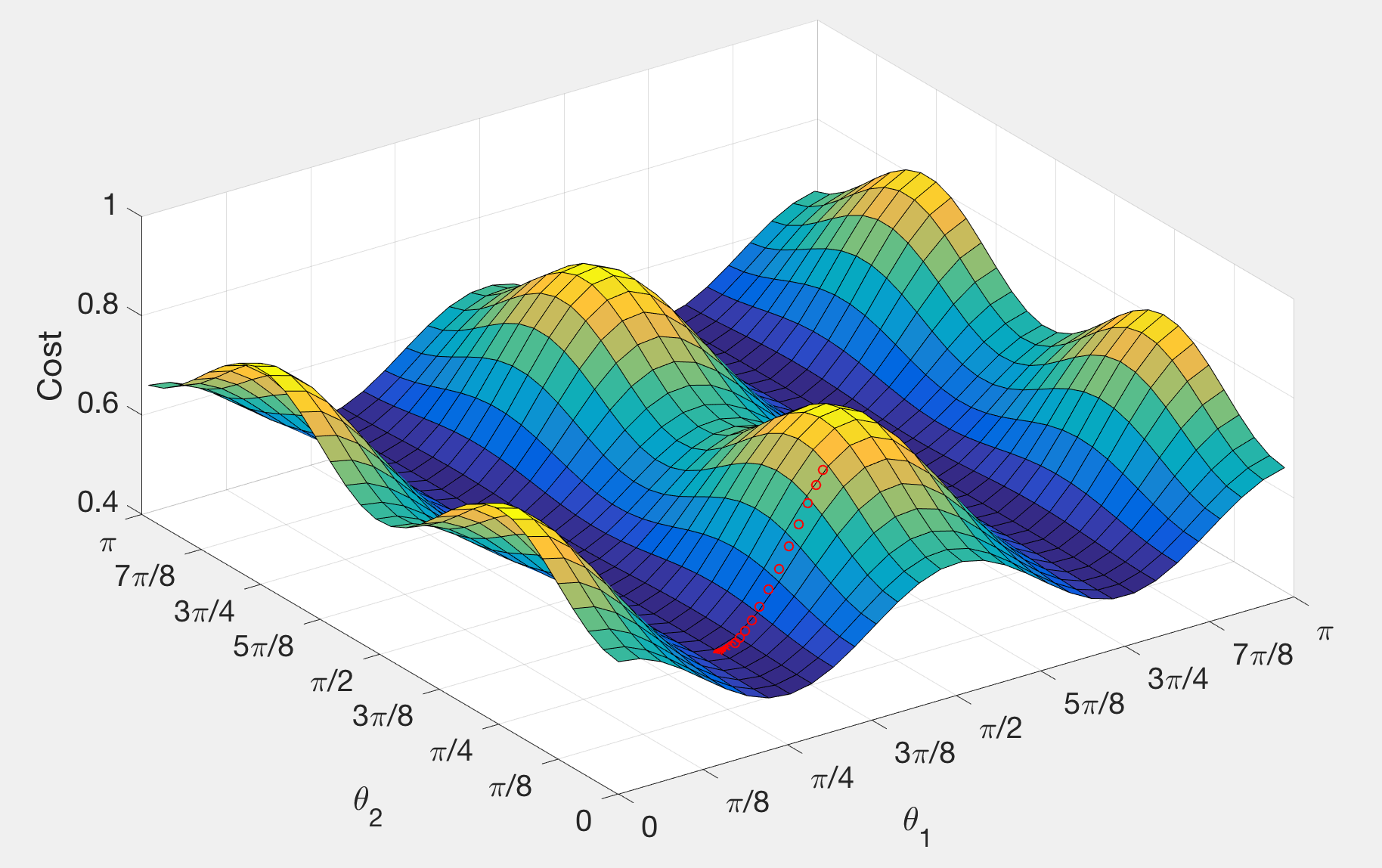}
\caption{\label{fig:cost_landscape_recovering_simon_circuit} This is
  the cost landscape sampled under a constant post processing
  permutation matrix. It can be shown that Simon's circuit lies in
a local minima via the red line which represented the gradient descent
path with a starting point close to Simon's quantum circuit solution
parameterised with $\theta_{1}$ and $\theta_{2}$.}
\end{figure}

\begin{figure}[ht]
\includegraphics[width=\linewidth]{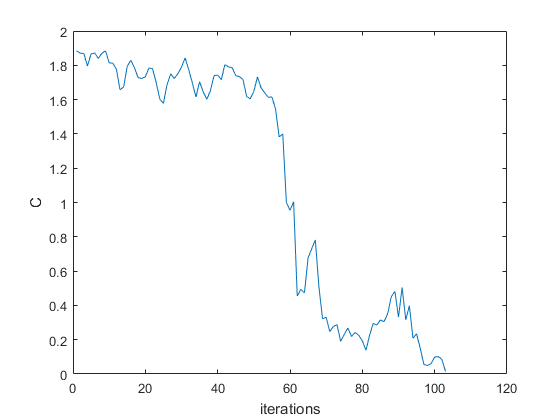}
\caption{\label{fig:C-T3} An example of the cost function value as a function of time during training. In this case, $n=3$ and there are 3 free parameters, like those free parameters in Eq.~\ref{eq:restricted_unitary}.}
\end{figure}

Fig.~\ref{fig:cost_landscape_recovering_simon_circuit} illustrates how Simon's solution $(\theta_{1}, \theta_{2}) =(\pi/4, \pi/4)$, is a local minimum of the cost function (for n=2). To check if this is truly optimal we did a brute force search over other circuits in this family and confirmed that they never achieve a lower cost value than Simon's solution. Fig.~\ref{fig:C-T3} shows, for $n = 3$, parameters staring at a random initial point, can also find the minimum of the cost function, which means can recover Simon's algorithm. To speed up the classic simulation on computer or high performance computer, we use the parallel computing toolbox in MATLAB, which supports parallel for-loops for running task-parallel algorithms on multiple processors and make the full use of multicore processors on the desktop via workers that run locally.\\

\subsection{Do not need all secret bit strings to train it}
We find that we can recover Simon's algorithm through training with just 1 secret string example for n=2, and 1 secret string for n=3. This is important as there are $2^n-1$ such secret strings (the null string is not counted, as it corresponds to a permutation rather than 2-1 function), as discussed earlier.
An initial approach, wherein the network would be asked to guess s after just one call to the oracle,  required all s's for the training.  
   
\section{Summary and Outlook}
 We trained a unitary network to find the optimal circuit for solving the task of finding the 
hidden bit string associated with Simon's oracle. The result is indeed that the circuit associated with Simon's algorithm is recovered, such that 
quantum parallelism is used to probe the oracle unitary. This demonstrates the potential of these techniques for finding algorithms.

We chose Simon's algorithm as a clean example of the hidden subgroup problem algorithms. It is plausible that the same approach can 
be used for other problems in that class. A key challenge is to find a task that is technologically useful, such as factoring, but which does not 
yet have a known quantum algorithm.
The current approach combines a human and machine to find the algorithm and making the machine discovery more autonomous may increase the chance of discovering algorithms we have not yet thought of.  

{\bf Note added:} Whilst we were preparing this manuscript, a paper with related ideas, for the case of Grover's algorithm, appeared on the pre-print server: arXiv:1805.09337, {\em Variationally learning Grover's Search Algorithm, Morales et al.}

\section*{Data availability statement}
\vspace{0.07cm}
This is a theoretical paper and there is no experimental data available beyond the numerical simulation data described in the paper. 

\section*{Acknowledgements}
Part of this work forms KHW's BSc thesis at Imperial College~\cite{Wan17}. We acknowledge discussions with Jon Allcock,  Heliang Huang, Sania Jevtic and Doug Plato. KHW is funded by the President's PhD Scholarship of Imperial College London. OD acknowledges funding from the 1000 Talents Youth project, China and the London Institute for Mathematical Sciences. FL acknowledges use of the SUSTech supercomputer Shuguang 6000. 
MK acknowledges funding from the Royal Society, a  Samsung GRO Grant, and a programme grant from the UK EPSRC (EP/K034480/1).

\bibliography{mybibfile.bib}

\begin{thebibliography}{32}%
\makeatletter
\providecommand \@ifxundefined [1]{%
 \@ifx{#1\undefined}
}%
\providecommand \@ifnum [1]{%
 \ifnum #1\expandafter \@firstoftwo
 \else \expandafter \@secondoftwo
 \fi
}%
\providecommand \@ifx [1]{%
 \ifx #1\expandafter \@firstoftwo
 \else \expandafter \@secondoftwo
 \fi
}%
\providecommand \natexlab [1]{#1}%
\providecommand \enquote  [1]{``#1''}%
\providecommand \bibnamefont  [1]{#1}%
\providecommand \bibfnamefont [1]{#1}%
\providecommand \citenamefont [1]{#1}%
\providecommand \href@noop [0]{\@secondoftwo}%
\providecommand \href [0]{\begingroup \@sanitize@url \@href}%
\providecommand \@href[1]{\@@startlink{#1}\@@href}%
\providecommand \@@href[1]{\endgroup#1\@@endlink}%
\providecommand \@sanitize@url [0]{\catcode `\\12\catcode `\$12\catcode
  `\&12\catcode `\#12\catcode `\^12\catcode `\_12\catcode `\%12\relax}%
\providecommand \@@startlink[1]{}%
\providecommand \@@endlink[0]{}%
\providecommand \url  [0]{\begingroup\@sanitize@url \@url }%
\providecommand \@url [1]{\endgroup\@href {#1}{\urlprefix }}%
\providecommand \urlprefix  [0]{URL }%
\providecommand \Eprint [0]{\href }%
\providecommand \doibase [0]{http://dx.doi.org/}%
\providecommand \selectlanguage [0]{\@gobble}%
\providecommand \bibinfo  [0]{\@secondoftwo}%
\providecommand \bibfield  [0]{\@secondoftwo}%
\providecommand \translation [1]{[#1]}%
\providecommand \BibitemOpen [0]{}%
\providecommand \bibitemStop [0]{}%
\providecommand \bibitemNoStop [0]{.\EOS\space}%
\providecommand \EOS [0]{\spacefactor3000\relax}%
\providecommand \BibitemShut  [1]{\csname bibitem#1\endcsname}%
\let\auto@bib@innerbib\@empty
\bibitem [{\citenamefont {Simon}(1997)}]{simonoriginalpaper}%
  \BibitemOpen
  \bibfield  {author} {\bibinfo {author} {\bibfnamefont {Daniel~R.}\
  \bibnamefont {Simon}},\ }\bibfield  {title} {\enquote {\bibinfo {title} {On
  the power of quantum computation},}\ }\href {\doibase
  10.11373/S0097539796298637} {\bibfield  {journal} {\bibinfo  {journal} {SIAM
  Journal on Computing}\ }\textbf {\bibinfo {volume} {26 (5)}},\ \bibinfo
  {pages} {1474 -- 1483} (\bibinfo {year} {1997})}\BibitemShut {NoStop}%
\bibitem [{\citenamefont {Shor}(1997)}]{Shor:1997:PAP:264393.264406}%
  \BibitemOpen
  \bibfield  {author} {\bibinfo {author} {\bibfnamefont {Peter~W.}\
  \bibnamefont {Shor}},\ }\bibfield  {title} {\enquote {\bibinfo {title}
  {Polynomial-time algorithms for prime factorization and discrete logarithms
  on a quantum computer},}\ }\href {\doibase 10.1137/S0097539795293172}
  {\bibfield  {journal} {\bibinfo  {journal} {SIAM J. Comput.}\ }\textbf
  {\bibinfo {volume} {26}},\ \bibinfo {pages} {1484--1509} (\bibinfo {year}
  {1997})}\BibitemShut {NoStop}%
\bibitem [{\citenamefont {Mermin}(2007)}]{mermin2007quantum}%
  \BibitemOpen
  \bibfield  {author} {\bibinfo {author} {\bibfnamefont {N.D.}\ \bibnamefont
  {Mermin}},\ }\href@noop {} {\emph {\bibinfo {title} {Quantum Computer
  Science: An Introduction}}}\ (\bibinfo  {publisher} {Cambridge University
  Press},\ \bibinfo {year} {2007})\BibitemShut {NoStop}%
\bibitem [{\citenamefont {Grigni}\ \emph {et~al.}(2001)\citenamefont {Grigni},
  \citenamefont {Schulman}, \citenamefont {Vazirani},\ and\ \citenamefont
  {Vazirani}}]{Grigni:2001:QMA:380752.380769}%
  \BibitemOpen
  \bibfield  {author} {\bibinfo {author} {\bibfnamefont {Michelangelo}\
  \bibnamefont {Grigni}}, \bibinfo {author} {\bibfnamefont {Leonard}\
  \bibnamefont {Schulman}}, \bibinfo {author} {\bibfnamefont {Monica}\
  \bibnamefont {Vazirani}}, \ and\ \bibinfo {author} {\bibfnamefont {Umesh}\
  \bibnamefont {Vazirani}},\ }\bibfield  {title} {\enquote {\bibinfo {title}
  {Quantum mechanical algorithms for the nonabelian hidden subgroup problem},}\
  }in\ \href {\doibase 10.1145/380752.380769} {\emph {\bibinfo {booktitle}
  {Proceedings of the Thirty-third Annual ACM Symposium on Theory of
  Computing}}},\ \bibinfo {series and number} {STOC '01}\ (\bibinfo
  {publisher} {ACM},\ \bibinfo {address} {New York, NY, USA},\ \bibinfo {year}
  {2001})\ pp.\ \bibinfo {pages} {68--74}\BibitemShut {NoStop}%
\bibitem [{\citenamefont {{Biamonte et.al.}}(2016)}]{Biamonte}%
  \BibitemOpen
  \bibfield  {author} {\bibinfo {author} {\bibnamefont {{Biamonte et.al.}}},\
  }\bibfield  {title} {\enquote {\bibinfo {title} {{Quantum Machine
  Learning}},}\ }\href@noop {} {\  (\bibinfo {year} {2016})},\ \Eprint
  {http://arxiv.org/abs/1607.08535} {arXiv:1607.08535} \BibitemShut {NoStop}%
\bibitem [{\citenamefont {Schuld}\ \emph {et~al.}(2014)\citenamefont {Schuld},
  \citenamefont {Sinayskiy},\ and\ \citenamefont {Petruccione}}]{Schuld14}%
  \BibitemOpen
  \bibfield  {author} {\bibinfo {author} {\bibfnamefont {M.}~\bibnamefont
  {Schuld}}, \bibinfo {author} {\bibfnamefont {I.}~\bibnamefont {Sinayskiy}}, \
  and\ \bibinfo {author} {\bibfnamefont {F.}~\bibnamefont {Petruccione}},\
  }\bibfield  {title} {\enquote {\bibinfo {title} {The quest for a quantum
  neural network},}\ }\href@noop {} {\bibfield  {journal} {\bibinfo  {journal}
  {Quantum Information Processing}\ }\textbf {\bibinfo {volume} {13}},\
  \bibinfo {pages} {2567--2586} (\bibinfo {year} {2014})}\BibitemShut {NoStop}%
\bibitem [{\citenamefont {Lloyd}\ \emph {et~al.}(2013)\citenamefont {Lloyd},
  \citenamefont {Mohseni},\ and\ \citenamefont {Rebentrost}}]{LloydMR13}%
  \BibitemOpen
  \bibfield  {author} {\bibinfo {author} {\bibfnamefont {S.}~\bibnamefont
  {Lloyd}}, \bibinfo {author} {\bibfnamefont {M.}~\bibnamefont {Mohseni}}, \
  and\ \bibinfo {author} {\bibfnamefont {P.}~\bibnamefont {Rebentrost}},\
  }\bibfield  {title} {\enquote {\bibinfo {title} {Quantum algorithms for
  supervised and unsupervised machine learning},}\ }\href@noop {} {\  (\bibinfo
  {year} {2013})},\ \Eprint {http://arxiv.org/abs/1307.0411} {arXiv:1307.0411
  [quant-ph]} \BibitemShut {NoStop}%
\bibitem [{\citenamefont {Lloyd}\ \emph {et~al.}(2014)\citenamefont {Lloyd},
  \citenamefont {Mohseni},\ and\ \citenamefont {Rebentrost}}]{LloydMR13ii}%
  \BibitemOpen
  \bibfield  {author} {\bibinfo {author} {\bibfnamefont {S.}~\bibnamefont
  {Lloyd}}, \bibinfo {author} {\bibfnamefont {M.}~\bibnamefont {Mohseni}}, \
  and\ \bibinfo {author} {\bibfnamefont {P.}~\bibnamefont {Rebentrost}},\
  }\bibfield  {title} {\enquote {\bibinfo {title} {Quantum principal component
  analysis},}\ }\href@noop {} {\bibfield  {journal} {\bibinfo  {journal}
  {Nature Physics}\ }\textbf {\bibinfo {volume} {10}},\ \bibinfo {pages}
  {631--633} (\bibinfo {year} {2014})}\BibitemShut {NoStop}%
\bibitem [{\citenamefont {Montanaro}(2015)}]{Montanaro15}%
  \BibitemOpen
  \bibfield  {author} {\bibinfo {author} {\bibfnamefont {A.}~\bibnamefont
  {Montanaro}},\ }\bibfield  {title} {\enquote {\bibinfo {title} {Quantum
  pattern matching fast on average},}\ }\href {\doibase
  10.1007/s00453-015-0060-4} {\bibfield  {journal} {\bibinfo  {journal}
  {Algorithmica}\ ,\ \bibinfo {pages} {1--24}} (\bibinfo {year}
  {2015})}\BibitemShut {NoStop}%
\bibitem [{\citenamefont {Aaronson}(2015)}]{Aaronson15}%
  \BibitemOpen
  \bibfield  {author} {\bibinfo {author} {\bibfnamefont {S.}~\bibnamefont
  {Aaronson}},\ }\bibfield  {title} {\enquote {\bibinfo {title} {Read the fine
  print},}\ }\href {http://dx.doi.org/10.1038/nphys3272} {\bibfield  {journal}
  {\bibinfo  {journal} {Nature Physics}\ }\textbf {\bibinfo {volume} {11}},\
  \bibinfo {pages} {291--293} (\bibinfo {year} {2015})}\BibitemShut {NoStop}%
\bibitem [{\citenamefont {Garnerone}\ \emph {et~al.}(2012)\citenamefont
  {Garnerone}, \citenamefont {Zanardi},\ and\ \citenamefont
  {Lidar}}]{GarneroneZL12}%
  \BibitemOpen
  \bibfield  {author} {\bibinfo {author} {\bibfnamefont {S.}~\bibnamefont
  {Garnerone}}, \bibinfo {author} {\bibfnamefont {P.}~\bibnamefont {Zanardi}},
  \ and\ \bibinfo {author} {\bibfnamefont {D.~A.}\ \bibnamefont {Lidar}},\
  }\bibfield  {title} {\enquote {\bibinfo {title} {Adiabatic quantum algorithm
  for search engine ranking},}\ }\href {\doibase
  10.1103/PhysRevLett.108.230506} {\bibfield  {journal} {\bibinfo  {journal}
  {Phys. Rev. Lett.}\ }\textbf {\bibinfo {volume} {108}},\ \bibinfo {pages}
  {230506} (\bibinfo {year} {2012})}\BibitemShut {NoStop}%
\bibitem [{\citenamefont {Harrow}\ \emph {et~al.}(2009)\citenamefont {Harrow},
  \citenamefont {Hassidim},\ and\ \citenamefont {Lloyd}}]{HarrowHL09}%
  \BibitemOpen
  \bibfield  {author} {\bibinfo {author} {\bibfnamefont {A.~W.}\ \bibnamefont
  {Harrow}}, \bibinfo {author} {\bibfnamefont {A.}~\bibnamefont {Hassidim}}, \
  and\ \bibinfo {author} {\bibfnamefont {S.}~\bibnamefont {Lloyd}},\ }\bibfield
   {title} {\enquote {\bibinfo {title} {Quantum algorithm for linear systems of
  equations},}\ }\href {\doibase 10.1103/PhysRevLett.103.150502} {\bibfield
  {journal} {\bibinfo  {journal} {Phys. Rev. Lett.}\ }\textbf {\bibinfo
  {volume} {103}},\ \bibinfo {pages} {150502} (\bibinfo {year}
  {2009})}\BibitemShut {NoStop}%
\bibitem [{\citenamefont {Lloyd}\ \emph {et~al.}(2016)\citenamefont {Lloyd},
  \citenamefont {Garnerone},\ and\ \citenamefont {Zanardi}}]{LloydGZ16}%
  \BibitemOpen
  \bibfield  {author} {\bibinfo {author} {\bibfnamefont {S.}~\bibnamefont
  {Lloyd}}, \bibinfo {author} {\bibfnamefont {S.}~\bibnamefont {Garnerone}}, \
  and\ \bibinfo {author} {\bibfnamefont {P.}~\bibnamefont {Zanardi}},\
  }\bibfield  {title} {\enquote {\bibinfo {title} {Quantum algorithms for
  topological and geometric analysis of big data},}\ }\href {\doibase
  10.1038/ncomms10138} {\bibfield  {journal} {\bibinfo  {journal} {Nature
  Communications}\ }\textbf {\bibinfo {volume} {7}},\ \bibinfo {pages} {10138}
  (\bibinfo {year} {2016})}\BibitemShut {NoStop}%
\bibitem [{\citenamefont {Rebentrost}\ \emph {et~al.}(2014)\citenamefont
  {Rebentrost}, \citenamefont {Mohseni},\ and\ \citenamefont
  {Lloyd}}]{RebenstrostML13}%
  \BibitemOpen
  \bibfield  {author} {\bibinfo {author} {\bibfnamefont {P.}~\bibnamefont
  {Rebentrost}}, \bibinfo {author} {\bibfnamefont {M.}~\bibnamefont {Mohseni}},
  \ and\ \bibinfo {author} {\bibfnamefont {S.}~\bibnamefont {Lloyd}},\
  }\bibfield  {title} {\enquote {\bibinfo {title} {Quantum support vector
  machine for big data classification},}\ }\href {\doibase
  10.1103/PhysRevLett.113.130503} {\bibfield  {journal} {\bibinfo  {journal}
  {Phys. Rev. Lett.}\ }\textbf {\bibinfo {volume} {113}},\ \bibinfo {pages}
  {130503} (\bibinfo {year} {2014})}\BibitemShut {NoStop}%
\bibitem [{\citenamefont {Wiebe}\ \emph {et~al.}(2012)\citenamefont {Wiebe},
  \citenamefont {Braun},\ and\ \citenamefont {Lloyd}}]{WiebeBL12}%
  \BibitemOpen
  \bibfield  {author} {\bibinfo {author} {\bibfnamefont {N.}~\bibnamefont
  {Wiebe}}, \bibinfo {author} {\bibfnamefont {D.}~\bibnamefont {Braun}}, \ and\
  \bibinfo {author} {\bibfnamefont {S.}~\bibnamefont {Lloyd}},\ }\bibfield
  {title} {\enquote {\bibinfo {title} {Quantum algorithm for data fitting},}\
  }\href {\doibase 10.1103/PhysRevLett.109.050505} {\bibfield  {journal}
  {\bibinfo  {journal} {Phys. Rev. Lett.}\ }\textbf {\bibinfo {volume} {109}},\
  \bibinfo {pages} {050505} (\bibinfo {year} {2012})}\BibitemShut {NoStop}%
\bibitem [{\citenamefont {Adcock}\ \emph {et~al.}(2015)\citenamefont {Adcock},
  \citenamefont {Allen}, \citenamefont {Day}, \citenamefont {Frick},
  \citenamefont {Hinchliff}, \citenamefont {Johnson}, \citenamefont
  {Morley-Short}, \citenamefont {Pallister}, \citenamefont {Price},\ and\
  \citenamefont {Stanisic}}]{Adcock15}%
  \BibitemOpen
  \bibfield  {author} {\bibinfo {author} {\bibfnamefont {J.}~\bibnamefont
  {Adcock}}, \bibinfo {author} {\bibfnamefont {E.}~\bibnamefont {Allen}},
  \bibinfo {author} {\bibfnamefont {M.}~\bibnamefont {Day}}, \bibinfo {author}
  {\bibfnamefont {S.}~\bibnamefont {Frick}}, \bibinfo {author} {\bibfnamefont
  {J.}~\bibnamefont {Hinchliff}}, \bibinfo {author} {\bibfnamefont
  {M.}~\bibnamefont {Johnson}}, \bibinfo {author} {\bibfnamefont
  {S.}~\bibnamefont {Morley-Short}}, \bibinfo {author} {\bibfnamefont
  {S.}~\bibnamefont {Pallister}}, \bibinfo {author} {\bibfnamefont
  {A.}~\bibnamefont {Price}}, \ and\ \bibinfo {author} {\bibfnamefont
  {S.}~\bibnamefont {Stanisic}},\ }\bibfield  {title} {\enquote {\bibinfo
  {title} {Advances in quantum machine learning},}\ }\href@noop {} {\
  (\bibinfo {year} {2015})},\ \Eprint {http://arxiv.org/abs/1512.02900}
  {arXiv:1512.02900 [quant-ph]} \BibitemShut {NoStop}%
\bibitem [{\citenamefont {Heim}\ \emph {et~al.}(2015)\citenamefont {Heim},
  \citenamefont {R{\o}nnow}, \citenamefont {Isakov},\ and\ \citenamefont
  {Troyer}}]{HeimRIT15}%
  \BibitemOpen
  \bibfield  {author} {\bibinfo {author} {\bibfnamefont {B.}~\bibnamefont
  {Heim}}, \bibinfo {author} {\bibfnamefont {T.~F.}\ \bibnamefont {R{\o}nnow}},
  \bibinfo {author} {\bibfnamefont {S.~V.}\ \bibnamefont {Isakov}}, \ and\
  \bibinfo {author} {\bibfnamefont {M.}~\bibnamefont {Troyer}},\ }\bibfield
  {title} {\enquote {\bibinfo {title} {Quantum versus classical annealing of
  {I}sing spin glasses},}\ }\href {\doibase 10.1126/science.aaa4170} {\bibfield
   {journal} {\bibinfo  {journal} {Science}\ }\textbf {\bibinfo {volume}
  {348}},\ \bibinfo {pages} {215--217} (\bibinfo {year} {2015})}\BibitemShut
  {NoStop}%
\bibitem [{\citenamefont {Gross}\ \emph {et~al.}(2010)\citenamefont {Gross},
  \citenamefont {Liu}, \citenamefont {Flammia}, \citenamefont {Becker},\ and\
  \citenamefont {Eisert}}]{GrossYFBE10}%
  \BibitemOpen
  \bibfield  {author} {\bibinfo {author} {\bibfnamefont {D.}~\bibnamefont
  {Gross}}, \bibinfo {author} {\bibfnamefont {Y.K.}\ \bibnamefont {Liu}},
  \bibinfo {author} {\bibfnamefont {S.~T.}\ \bibnamefont {Flammia}}, \bibinfo
  {author} {\bibfnamefont {S.}~\bibnamefont {Becker}}, \ and\ \bibinfo {author}
  {\bibfnamefont {J.}~\bibnamefont {Eisert}},\ }\bibfield  {title} {\enquote
  {\bibinfo {title} {Quantum state tomography via compressed sensing},}\ }\href
  {\doibase 10.1103/PhysRevLett.105.150401} {\bibfield  {journal} {\bibinfo
  {journal} {Phys. Rev. Lett.}\ }\textbf {\bibinfo {volume} {105}},\ \bibinfo
  {pages} {150401} (\bibinfo {year} {2010})}\BibitemShut {NoStop}%
\bibitem [{\citenamefont {Dunjko}\ \emph {et~al.}(2016)\citenamefont {Dunjko},
  \citenamefont {Taylor},\ and\ \citenamefont {Briegel}}]{Dunjko16}%
  \BibitemOpen
  \bibfield  {author} {\bibinfo {author} {\bibfnamefont {V.}~\bibnamefont
  {Dunjko}}, \bibinfo {author} {\bibfnamefont {J.~M.}\ \bibnamefont {Taylor}},
  \ and\ \bibinfo {author} {\bibfnamefont {H.~J.}\ \bibnamefont {Briegel}},\
  }\bibfield  {title} {\enquote {\bibinfo {title} {Quantum-enhanced machine
  learning},}\ }\href {\doibase 10.1103/PhysRevLett.117.130501} {\bibfield
  {journal} {\bibinfo  {journal} {Phys. Rev. Lett.}\ }\textbf {\bibinfo
  {volume} {117}},\ \bibinfo {pages} {130501} (\bibinfo {year}
  {2016})}\BibitemShut {NoStop}%
\bibitem [{\citenamefont {Wittek}(2014)}]{Wittek14}%
  \BibitemOpen
  \bibinfo {editor} {\bibfnamefont {P.}~\bibnamefont {Wittek}},\ ed.,\ \href
  {\doibase http://dx.doi.org/10.1016/B978-0-12-800953-6.00015-3} {\emph
  {\bibinfo {title} {Quantum Machine Learning}}}\ (\bibinfo  {publisher}
  {Academic Press},\ \bibinfo {address} {Boston},\ \bibinfo {year}
  {2014})\BibitemShut {NoStop}%
\bibitem [{\citenamefont {Bisio}\ \emph {et~al.}(2010)\citenamefont {Bisio},
  \citenamefont {Chiribella}, \citenamefont {D'Ariano}, \citenamefont
  {Facchini},\ and\ \citenamefont {Perinotti}}]{Bisio10}%
  \BibitemOpen
  \bibfield  {author} {\bibinfo {author} {\bibfnamefont {A.}~\bibnamefont
  {Bisio}}, \bibinfo {author} {\bibfnamefont {G.}~\bibnamefont {Chiribella}},
  \bibinfo {author} {\bibfnamefont {G.~M.}\ \bibnamefont {D'Ariano}}, \bibinfo
  {author} {\bibfnamefont {S.}~\bibnamefont {Facchini}}, \ and\ \bibinfo
  {author} {\bibfnamefont {P.}~\bibnamefont {Perinotti}},\ }\bibfield  {title}
  {\enquote {\bibinfo {title} {Optimal quantum learning of a unitary
  transformation},}\ }\href {\doibase 10.1103/PhysRevA.81.032324} {\bibfield
  {journal} {\bibinfo  {journal} {Phys. Rev. A}\ }\textbf {\bibinfo {volume}
  {81}},\ \bibinfo {pages} {032324} (\bibinfo {year} {2010})}\BibitemShut
  {NoStop}%
\bibitem [{\citenamefont {Sasaki}\ and\ \citenamefont
  {Carlini}(2002)}]{Sasaki02}%
  \BibitemOpen
  \bibfield  {author} {\bibinfo {author} {\bibfnamefont {M.}~\bibnamefont
  {Sasaki}}\ and\ \bibinfo {author} {\bibfnamefont {A.}~\bibnamefont
  {Carlini}},\ }\bibfield  {title} {\enquote {\bibinfo {title} {Quantum
  learning and universal quantum matching machine},}\ }\href {\doibase
  10.1103/PhysRevA.66.022303} {\bibfield  {journal} {\bibinfo  {journal} {Phys.
  Rev. A}\ }\textbf {\bibinfo {volume} {66}},\ \bibinfo {pages} {022303}
  (\bibinfo {year} {2002})}\BibitemShut {NoStop}%
\bibitem [{\citenamefont {{Bang}}\ \emph {et~al.}(2008)\citenamefont {{Bang}},
  \citenamefont {{Lim}}, \citenamefont {{Kim}},\ and\ \citenamefont
  {{Lee}}}]{BangLKL08}%
  \BibitemOpen
  \bibfield  {author} {\bibinfo {author} {\bibfnamefont {J.}~\bibnamefont
  {{Bang}}}, \bibinfo {author} {\bibfnamefont {J.}~\bibnamefont {{Lim}}},
  \bibinfo {author} {\bibfnamefont {M.~S.}\ \bibnamefont {{Kim}}}, \ and\
  \bibinfo {author} {\bibfnamefont {J.}~\bibnamefont {{Lee}}},\ }\bibfield
  {title} {\enquote {\bibinfo {title} {{Quantum Learning Machine}},}\
  }\href@noop {} {\bibfield  {journal} {\bibinfo  {journal} {ArXiv e-prints}\ }
  (\bibinfo {year} {2008})},\ \Eprint {http://arxiv.org/abs/0803.2976}
  {arXiv:0803.2976 [quant-ph]} \BibitemShut {NoStop}%
\bibitem [{\citenamefont {Sent{\'i}s}\ \emph {et~al.}(2015)\citenamefont
  {Sent{\'i}s}, \citenamefont {Gu{\c{T}}{\u{a}}},\ and\ \citenamefont
  {Adesso}}]{Sentis15}%
  \BibitemOpen
  \bibfield  {author} {\bibinfo {author} {\bibfnamefont {G.}~\bibnamefont
  {Sent{\'i}s}}, \bibinfo {author} {\bibfnamefont {M.}~\bibnamefont
  {Gu{\c{T}}{\u{a}}}}, \ and\ \bibinfo {author} {\bibfnamefont
  {G.}~\bibnamefont {Adesso}},\ }\bibfield  {title} {\enquote {\bibinfo {title}
  {Quantum learning of coherent states},}\ }\href {\doibase
  10.1140/epjqt/s40507-015-0030-4} {\bibfield  {journal} {\bibinfo  {journal}
  {EPJ Quantum Technology}\ }\textbf {\bibinfo {volume} {2}},\ \bibinfo {pages}
  {17} (\bibinfo {year} {2015})}\BibitemShut {NoStop}%
\bibitem [{\citenamefont {Banchi}\ \emph {et~al.}(2016)\citenamefont {Banchi},
  \citenamefont {Pancotti},\ and\ \citenamefont {Bose}}]{Banchi16}%
  \BibitemOpen
  \bibfield  {author} {\bibinfo {author} {\bibfnamefont {L.}~\bibnamefont
  {Banchi}}, \bibinfo {author} {\bibfnamefont {N.}~\bibnamefont {Pancotti}}, \
  and\ \bibinfo {author} {\bibfnamefont {S.}~\bibnamefont {Bose}},\ }\bibfield
  {title} {\enquote {\bibinfo {title} {Quantum gate learning in qubit networks:
  Toffoli gate without time-dependent control},}\ }\href@noop {} {\bibfield
  {journal} {\bibinfo  {journal} {Npj Quantum Information}\ }\textbf {\bibinfo
  {volume} {2}},\ \bibinfo {pages} {16019 EP --} (\bibinfo {year}
  {2016})}\BibitemShut {NoStop}%
\bibitem [{\citenamefont {Palittapongarnpim}\ \emph {et~al.}(2016)\citenamefont
  {Palittapongarnpim}, \citenamefont {Wittek}, \citenamefont {Zahedinejad},
  \citenamefont {Vedaie},\ and\ \citenamefont {Sanders}}]{Palittapongarnpim16}%
  \BibitemOpen
  \bibfield  {author} {\bibinfo {author} {\bibfnamefont {P.}~\bibnamefont
  {Palittapongarnpim}}, \bibinfo {author} {\bibfnamefont {P.}~\bibnamefont
  {Wittek}}, \bibinfo {author} {\bibfnamefont {E.}~\bibnamefont {Zahedinejad}},
  \bibinfo {author} {\bibfnamefont {S}~\bibnamefont {Vedaie}}, \ and\ \bibinfo
  {author} {\bibfnamefont {B.~C.}\ \bibnamefont {Sanders}},\ }\bibfield
  {title} {\enquote {\bibinfo {title} {Learning in quantum control:
  high-dimensional global optimization for noisy quantum dynamics},}\
  }\href@noop {} {\  (\bibinfo {year} {2016})}\BibitemShut {NoStop}%
\bibitem [{\citenamefont {{Wan}}\ \emph {et~al.}(2017)\citenamefont {{Wan}},
  \citenamefont {{Dahlsten}}, \citenamefont {{Kristj{\'a}nsson}}, \citenamefont
  {{Gardner}},\ and\ \citenamefont {{Kim}}}]{WanDKGK17}%
  \BibitemOpen
  \bibfield  {author} {\bibinfo {author} {\bibfnamefont {K.~H.}\ \bibnamefont
  {{Wan}}}, \bibinfo {author} {\bibfnamefont {O.}~\bibnamefont {{Dahlsten}}},
  \bibinfo {author} {\bibfnamefont {H.}~\bibnamefont {{Kristj{\'a}nsson}}},
  \bibinfo {author} {\bibfnamefont {R.}~\bibnamefont {{Gardner}}}, \ and\
  \bibinfo {author} {\bibfnamefont {M.~S.}\ \bibnamefont {{Kim}}},\ }\bibfield
  {title} {\enquote {\bibinfo {title} {{Quantum generalisation of feedforward
  neural networks}},}\ }\href {\doibase 10.1038/s41534-017-0032-4} {\bibfield
  {journal} {\bibinfo  {journal} {npj Quantum Information}\ }\textbf {\bibinfo
  {volume} {3}},\ \bibinfo {eid} {36} (\bibinfo {year} {2017})}\BibitemShut
  {NoStop}%
\bibitem [{\citenamefont {Vazirani}(2004)}]{vaziranilec7}%
  \BibitemOpen
  \bibfield  {author} {\bibinfo {author} {\bibfnamefont {Umesh~V.}\
  \bibnamefont {Vazirani}},\ }\href
  {https://people.eecs.berkeley.edu/~vazirani/f04quantum/notes/lec7.pdf}
  {\enquote {\bibinfo {title} {Simon’s algorithm},}\ } (\bibinfo {year}
  {2004}),\ \bibinfo {note} {unpublished}\BibitemShut {NoStop}%
\bibitem [{\citenamefont {Coppersmith}\ and\ \citenamefont
  {Winograd}(1990)}]{cwalgorithm}%
  \BibitemOpen
  \bibfield  {author} {\bibinfo {author} {\bibfnamefont {Don}\ \bibnamefont
  {Coppersmith}}\ and\ \bibinfo {author} {\bibfnamefont {Shmuel}\ \bibnamefont
  {Winograd}},\ }\bibfield  {title} {\enquote {\bibinfo {title} {Matrix
  multiplication via arithmetic progressions},}\ }\href {\doibase
  10.1016/S0747-7171(08)80013-2} {\bibfield  {journal} {\bibinfo  {journal}
  {Journal of Symbolic Computation}\ }\textbf {\bibinfo {volume} {9 (3)}},\
  \bibinfo {pages} {251} (\bibinfo {year} {1990})}\BibitemShut {NoStop}%
\bibitem [{\citenamefont {Nielsen}(1991)}]{Nielsen_nn_online_bk}%
  \BibitemOpen
  \bibfield  {author} {\bibinfo {author} {\bibfnamefont {M.~A.}\ \bibnamefont
  {Nielsen}},\ }\href@noop {} {\emph {\bibinfo {title} {Neural Networks and
  Deep Learning}}}\ (\bibinfo  {publisher} {Determination Press},\ \bibinfo
  {address} {online book},\ \bibinfo {year} {1991})\BibitemShut {NoStop}%
\bibitem [{\citenamefont {de~Weck}(2010)}]{gene_1}%
  \BibitemOpen
  \bibfield  {author} {\bibinfo {author} {\bibfnamefont {Olivier}\ \bibnamefont
  {de~Weck}},\ }\href
  {https://ocw.mit.edu/courses/engineering-systems-division/esd-77-multidisciplinary-system-design-optimization-spring-2010/lecture-notes/MITESD_77S10_lec11.pdf}
  {\enquote {\bibinfo {title} {A basic introduction to genetic algorithms},}\ }
  (\bibinfo {year} {2010}),\ \bibinfo {note} {online notes,
  unpublished}\BibitemShut {NoStop}%
\bibitem [{\citenamefont {Wan}(2017)}]{Wan17}%
  \BibitemOpen
  \bibfield  {author} {\bibinfo {author} {\bibfnamefont {Kwok~Ho}\ \bibnamefont
  {Wan}},\ }\href@noop {} {\enquote {\bibinfo {title} {The power of quantum
  computation-a new perspective, {I}mperial {C}ollege {L}ondon},}\ } (\bibinfo
  {year} {2017}),\ \bibinfo {note} {{B}.{Sc}. Thesis}\BibitemShut {NoStop}%
\end{thebibliography}%

\end{document}